\documentstyle[12pt,a4]{article} 
\newcommand{\notC}{\mbox{{\sf C}\hspace*{-0.18cm}{\sf
l}\hspace*{0.1cm}}} \newcommand{\an}{\alpha} 
\begin{document}

\title{\vspace{-2cm} \hfill {\small THEF-NYM 94.2}\vspace{-.3cm} \\
\hfill {\small hep-th/9403041}\\ \vspace{-.3cm} \hfill {\small March
1994} \\ \vspace{1cm} {\Large \bf EXACT STRING BACKGROUND FROM A WZW
MODEL BASED ON THE HEISENBERG GROUP}} \vspace{2.5cm} \author{
{\normalsize \bf Alexandros A. Kehagias} \thanks{Partially supported by
CEC Contract No. ERBCHBGCT920197} \thanks{e-mail: kehagias@sci.kun.nl}
\\ [.1cm] {\normalsize and} \\ [.1cm]
  {\normalsize \bf Patrick A. A.  Meessen} \thanks{Partially supported
by `de Wet op Studiefinanciering' No. 5614-71105-0-13} \thanks{e-mail:
patrickm@sci.kun.nl} \\ [.2cm] \\ {\normalsize     Institute of
Theoretical Physics } \\ {\normalsize   University of Nijmegen,
Toernooiveld 1,} \\ {\normalsize    6525 ED, Nijmegen, The Netherlands
}\\ [.1cm]} \date{} \maketitle

\begin{abstract} \begin{sloppypar} \normalsize

A WZW model for the non-semi-simple $D$-dimensional Heisenberg group,
which directly generalizes the $E_{2}^{c}$ group, is constructed.  It
is found to correspond to an $D$-dimensional string background of
plane-wave type with physical Lorentz signature $D$-2.  Perturbative
and non-perturbative considerations lead both to an integer central
charge $c$=$D$.  \end{sloppypar} \end{abstract}

\vspace{3cm}

\noindent \newpage Recently, WZW models based on non-semi-simple groups
have attained a lot of interest [1--6], particularly because they lead
to exactly solvable models corresponding to exact string backgrounds.
To write down a WZW model for a non-semi-simple group is not
straightforward since its Killing metric is degenerate.  However, it is
possible that another, non-degenerate, invariant bilinear form on the
algebra exists which allows for a WZW model to be constructed.  The
non-semi-simple groups explicitly studied so far are the centrally
extended 2D Euclidean group $E_{2}^{c}$ \cite{1}, its d-dimensional
generalization $E_{d}^{c}$ \cite{3}, as well as a large class of
non-semi-simple groups arising after contraction of {\cal
G}$\times${\cal H} where {\cal H} is a subgroup of {\cal G} \cite{2}.
In particular, $E_{2}^c$ is isomorphic to the 4-dimensional Heisenberg
group.  It is then natural to discuss its $D$-dimensional analogue,
namely the Heisenberg group {\cal H}$_{D}$.  For this, a non-degenerate
bilinear form exists, a WZW model can be written down and a
Sugawara-type construction can be carried out.  The resulting
$D$-dimensional conformal $\sigma$-model has one null Killing vector
and the corresponding conformal field theory has $c=D$, in accordance
with the general result of \cite{4}.  \par
 The WZW models based on a group {\cal G} are defined on a two-surface
 $\Sigma$ by the action \begin{equation}
  {\cal S}_{WZW} \,=\, -\frac{k}{4 \pi} \int_{\Sigma} Tr(g^{-1}dg \,
  g^{-1}dg ) + \frac{k}{6 \pi} \int_{B} Tr(g^{-1}dg \wedge g^{-1}dg
  \wedge g^{-1}dg) \, , \label{w1} \end{equation} where $g^{-1}dg$ are
the right invariant 1-forms on {\cal G} and $B$ is a three manifold
bounded by $\Sigma$.  The right invariant 1-forms are elements of the
Lie algebra of {\cal G}, so that they may be expressed as
\begin{equation}
  g^{-1}dg \, = \, i A^{I} T_{I} \, , \label{w2} \end{equation} where
$T_{I} \; (I=1,\ldots,dimG)$ are the generators of the group satisfying
\begin{equation}
    [ T_{I} , T_{J} ] \, = \, i {f_{IJ}}^{K} T_{K} \, .\label{w3}
\end{equation} The WZW action (\ref{w1}) can then be written in terms
of the $A^{I}$'s as \begin{equation} {\cal S}_{WZW} \,=\, -\frac{k}{4
\pi} \int_{\Sigma} d^{2} \sigma A^{I}_{\alpha}A^{J\alpha} \Omega_{IJ}
\,-\, \frac{k}{12 \pi} \int_{B} d^{3}\sigma \, \epsilon^{\alpha \beta
\gamma} A^{I}_{\alpha} A^{J}_{\beta} A^{L}_{\gamma} {f_{IJ}}^{N}
\Omega_{NL}\, ,  \label{w4} \end{equation} where $\Omega_{IJ} =
Tr(T_{I}T_{J})$ is the Killing metric for the group {\cal G}.  However,
when the group is non-semi-simple, the Killing metric is degenerate and
one might replace it, as was proposed in \cite{1}, by a symmetric,
invariant and non-degenerate form $\Omega_{IJ}$ that satisfies
\begin{equation}
    {f_{IJ}}^{K}\Omega_{KL} \,+\, {f_{IL}}^{K}\Omega_{KJ} \, = \, 0 \,
    .\label{w5} \end{equation} Such a form can be found, apart from the
above mentioned cases, also for the Heisenberg group {\cal H}$_{D}$.
\par Before proceeding to the general case, let us first examine the
case {\cal H}$_{4}$, which originates from the dynamics of a single
one-dimensional harmonic oscillator.
 {\cal H}$_{4}$ is generated by $\{ \an$, $\an^{\dagger}$,
 $N=\an^{\dagger} \an$, $I\}$ and the commutation relations
\begin{eqnarray}
  [ \an , \an^{\dagger} ] \;&=&\; I  \nonumber \, ,\\{} [ N ,
  \an^{\dagger} ] \;&=&\; \an^{\dagger} \nonumber \, , \\{} [ N , \an ]
  \;&=&\; -\an \, .\label{w6} \end{eqnarray} It is non-semi-simple so
that its Killing form is degenerate.  A non-degenerate solution to
eq.(\ref{w5}) exists and is given by \begin{equation}
 \Omega_{IJ} \;=\; \left( \begin{array}{cccc} 0&a&0&0 \\ a&0&0&0
 \\ 0&0&b&-a \\ 0&0&-a&0 \end{array} \right) \, , \,  \Omega^{IJ} \;=\;
 \left( \begin{array}{cccc} 0&\frac{1}{a}&0&0 \\ \frac{1}{a}&0&0&0
 \\ 0&0&0&-\frac{1}{a} \\ 0&0&-\frac{1}{a}&-\frac{b}{a^{2}} \end{array}
\right) \,. \label{w7} \end{equation} \par A general element of {\cal
H}$_{4}$ is written as \begin{equation}
    g \;=\; e^{i q \an + i \bar{q} \an^{\dagger}} e^{i u N + i v I} \,
    ,\label{w8} \end{equation} where the first term on the right hand
side of eq.(\ref{w8}) is an element of the coset space {\cal
H}$_{4}/U(1)\! \times \! U(1) \sim \notC$ parametrized by the complex
coordinates $q, \bar{q}$.  By using the relations \begin{eqnarray}
   e^{iq\an + i\bar{q} \an^{\dagger}} \;&=&\;
   e^{iq\an}e^{i\bar{q}\an^{\dagger}}e^{\frac{q \bar{q}}{2}} \, ,
   \nonumber \\ e^{-i u N} \an e^{iuN} \;&=&\; e^{iu} \an \, ,
   \nonumber \\ e^{-i \bar{q} \an^{\dagger}} \an e^{i \bar{q}
   \an^{\dagger}} \;&=&\; \an \,+\, i\bar{q} I \, , \label{w9}
\end{eqnarray} we find that \begin{equation} g^{-1}dg \,=\, ie^{iu}dq
\an \,+\, ie^{-iu}d\bar{q}\an^{\dagger} \,+\, iduN \,+\, (idv +
\frac{1}{2}qd\bar{q} - \frac{1}{2} \bar{q}dq)I \, , \label{w10}
\end{equation} so that the $A^{I}$'s in eq.(\ref{w2}) are given by
\begin{eqnarray}
  A^{1} \;&=&\; e^{iu} dq \, ,  \nonumber \\ A^{2} \;&=&\; e^{-iu}
  d\bar{q} \, , \nonumber \\ A^3 \;&=&\; du \, , \nonumber \\ A^4
  \;&=&\; dv \,-\, \frac{i}{2}qd\bar{q} \,+\, \frac{i}{2}\bar{q}dq \,
  .\label{w11} \end{eqnarray} The terms that are being integrated over
in (\ref{w4}) are calculated to be \begin{eqnarray}
 A^{I}_{\alpha}A^{J\alpha}\Omega_{IJ} \;&=&\; 2a\partial_{\alpha}q
 \partial^{\alpha} \bar{q} -2a(\partial_{\alpha} v -\frac{i}{2}q
 \partial_{\alpha} \bar{q} +\frac{i}{2}\bar{q}\partial_{\alpha}q)
 \partial^{\alpha}u +b \partial_{\alpha}u \partial^{\alpha} u \, ,
 \nonumber \\ \epsilon^{\alpha \beta \gamma}
 A^{I}_{\alpha}A^{J}_{\beta}A^{K}_{\gamma} f_{IJK} \;&=&\; 6ia
 \epsilon_{\alpha \beta \gamma} \partial^{\gamma} (u\partial^{\alpha}q
 \partial^{\beta} \bar{q} ) \, , \label{w12} \end{eqnarray} and the WZW
action (\ref{w4}) is written as \begin{eqnarray} {\cal S}_{WZW} \,&=&
-\frac{k}{4 \pi} \int d^{2}\sigma \left( \rule{0mm}{.5cm}2a
\partial_{\alpha}q\partial^{\alpha}\bar{q}-2a(\partial_{\alpha}v-\frac{i}{2}q\partial_{\alpha}\bar{q}+\frac{i}{2}\bar{q}\partial_{\alpha}q)
\partial^{\alpha}u \right. \nonumber \\ &&+ \left. b \partial_{\alpha}u
\partial^{\alpha}u +2ia\epsilon_{\alpha \beta} u \partial^{\alpha}q
\partial^{\beta} \bar{q} \rule{0mm}{0.5cm} \right) \, . \label{w13}
\end{eqnarray} By regarding this action as a $\sigma$-model action of
the form \begin{equation}
 {\cal S} \;=\; \int d^{2} \sigma ( G_{\mu \nu} \partial_{\alpha}
 X^{\mu} \partial^{\alpha} X^{\nu} \,+\, B_{\mu \nu} \epsilon^{\alpha
 \beta} \partial_{\alpha} X^{\mu} \partial_{\beta} X^{\nu} )\, ,
 \label{w14} \end{equation} we can read off the background space-time
metric and the antisymmetric field.  In the coordinate base $[dq,
d\bar{q}, du,dv]$ they are given, up to multiplicative factors, by
\begin{eqnarray}
    G_{\mu \nu} \;&=&\; \left( \begin{array}{cccc} 0&\frac{1}{2}&
    -\frac{i}{4}\bar{q}&0 \\ \frac{1}{2}&0& \frac{i}{4}q&0
    \\ -\frac{i}{4}\bar{q}&\frac{i}{4}q&\beta^{2}&\frac{1}{2}
    \\ 0&0&\frac{1}{2}&0 \end{array} \right)  \, , \nonumber \\ B_{q
    \bar{q}} \;&=&\; \frac{i}{2}u \, , \label{w15} \end{eqnarray} where
$\beta^{2} = \frac{b}{2a}$, and thus, the background space-time line
element is given by \begin{equation}
 ds^{2} \;=\;
 dqd\bar{q}-(dv-\frac{i}{2}qd\bar{q}+\frac{i}{2}\bar{q}dq)du+\beta^{2}
 du^2 \, . \label{w16} \end{equation} It describes a plane-wave
space-time \cite{7,8} and it can be identified to the one found by
Nappi and Witten in \cite{1} by virtue of the transformation
$q=a_{1}-ia_{2}$ and $\bar{q}=a_{1}+ia_{2}$.  By introducing polar
coordinates $q=Re^{i\theta}$, $\bar{q}=Re^{-i\theta}$, the line element
turns out to be \begin{equation}
 ds^2 \;=\; dR^2 + R^{2}d\theta^2 -(dv-R^{2}d\theta)du + \beta^{2} du^2
 \, . \label{w17} \end{equation} The signature of this metric is
manifest in the orthonormal base \begin{eqnarray}
   e^0 \;&=&\; \frac{1}{2 \beta} (dv - R^{2}d\theta) \, , \nonumber \\
   e^1 \;&=&\; dR \, , \nonumber \\ e^2 \;&=&\; Rd\theta \, , \nonumber
   \\ e^3 \;&=&\; \beta du - e^{0} \, , \label{w18} \end{eqnarray}
where the metric is $\eta_{\mu \nu} = (-1,+1,+1,+1)$.  The only
non-vanishing components of the Ricci tensor in the above base are
\begin{equation}
   R_{00} \,=\, R_{33} \,=\, R_{03} \,=\, \frac{1}{2\beta^{2}} \, .
   \label{w19} \end{equation} In the same base, the antisymmetric two
form field $B= \frac{1}{2} B_{\mu \nu} dx^{\mu} \wedge dx^{\nu}$ is
written as \begin{equation}
	  B \;=\; u e^{1} \wedge e^{2} \, , \label{w20}
\end{equation} and thus \begin{equation}
   H \,=\, dB \,=\, \frac{1}{\beta} \, e^{3} \wedge e^{1} \wedge e^{2}
   +  \frac{1}{\beta} \, e^{0} \wedge e^{1} \wedge e^{2}  \, .
   \label{w21} \end{equation} The non-vanishing components of $H$ are
then \begin{equation}
    H_{123} \,=\, H_{012} \,=\, \frac{1}{\beta} \, . \label{w22}
\end{equation} Employing eqs.(\ref{w19}), (\ref{w22}) in the one-loop
beta-function equations \begin{eqnarray}
 0 \;&=&\;  -R + \frac{1}{12} H^{2} + 4 (\nabla \Phi )^{2} - 4
 \nabla^{2} \Phi \,+\, \frac{2(c-4)}{3 \alpha^{'}} \, , \nonumber \\ 0
 \;&=&\; R_{\mu \nu} -\frac{1}{4} H_{\mu \nu}^{2} +
 2\nabla_{\mu}\nabla_{\nu} \Phi \, , \nonumber \\ 0 \;&=&\;
 \nabla_{\lambda}H^{\lambda}_{\mu \nu} -2 (\nabla_{\lambda}\Phi)
 H^{\lambda}_{\mu \nu} \, , \label{w23} \end{eqnarray} one can find
that the dilaton is constant and that the central charge is four
($c=4$).  (A constant dilaton is expected on behalf of the homogeneity
of the space.) \par Let us now discuss the general case {\cal
H}$_{D}$.  The Heisenberg group {\cal H}$_{D}$ is a $D$-dimensional
group generated by the set of generators $\{ \an_{i},\,
\an_{i}^{\dagger},$ \mbox{$N=\sum_{i} \an_{i}^{\dagger} \an_{i}$}, $I,$
\mbox{$i=1,\ldots ,r=\frac{D-2}{2} \}$} and the commutation relations
\begin{eqnarray}
    [ \an_{i} , \an_{j}^{\dagger} ] \;&=&\; \delta_{ij} I \, ,
    \nonumber \\{} [ N , \an_{i} ] \;&=&\; -\an_{i} \, , \nonumber \\{}
    [ N , \an_{i}^{\dagger} ] \;&=&\; \an_{i}^{\dagger} \, , \nonumber
    \\{} [N, I] \;&=&\; [ \an_{i} , I] \;=\; [ \an_{i}^{\dagger} , I ]
    \;=\; 0 \, . \label{w24} \end{eqnarray} The Killing form of {\cal
H}$_{D}$ is degenerate and a non-degenerate solution to eq.(\ref{w5})
which can replace the Killing form is given by \begin{equation}
    \Omega_{IJ} \;=\; \left( \begin{array}{ccccc}
    A_{1}&&&&\\ &\ddots&&&\\ &&A_{r}&&\\ &&&b&-a\\ &&&-a&0 \end{array}
    \right) \, , \label{w24a} \end{equation} where \begin{equation}
  A_{1}\,=\, A_{2} \,= \ldots =\, A_{r} \,=\, \left( \begin{array}{cc}
  0&a\\ a&0 \end{array} \right) \, . \label{w24b} \end{equation} An
element of {\cal H}$_{D}$ is expressed as \begin{equation}
    g \;=\; e^{i\sum_{i} ( q_{i} \an_{i} + \bar{q}_{i}
    \an_{i}^{\dagger})} e^{iuN+ivI} \, , \label{w26} \end{equation} and
proceeding as before, we find that \begin{eqnarray}
   A^{i} \;&=&\; e^{iu} dq_{i} \, , \nonumber \\ \bar{A}^{i} \;&=&\;
   e^{-iu} d\bar{q}_{i} \, , \nonumber \\ A^{D-1} \;&=&\; du \, ,
   \nonumber \\ A^D \;&=&\; dv - \frac{i}{2} \sum_{i} (
   q_{i}d\bar{q}_{i}-\bar{q}_{i}dq_{i} ) \, . \label{w27}
\end{eqnarray} The WZW action is then written as \begin{eqnarray}
   {\cal S}_{WZW} \,&=&\, -\frac{k}{4 \pi} \int_{\Sigma} d^{2}\sigma
   \left( \rule{0mm}{.5cm}  2a \sum_{i} \partial_{\alpha}q_{i}
   \partial^{\alpha} \bar{q}_{i} -2a ( \partial_{\alpha}v -\frac{i}{2}
   \sum_{i} q_{i}\partial_{\alpha} \bar{q}_{i}- \bar{q}_{i}
   \partial_{\alpha} q_{i} ) \partial^{\alpha} u \right. \nonumber \\
  &+& \left. b \partial_{\alpha} u \partial^{\alpha} u
  +2ia\epsilon_{\alpha \beta} u \sum_{i} \partial^{\alpha}q_{i}
  \partial^{\beta} \bar{q}_{i} \rule{0mm}{.5cm} \right) \, .
  \label{w28} \end{eqnarray} By comparing this action with
eq.(\ref{w14}) we can read off the space-time metric, which is written,
up to multiplicative factors, as \par \begin{equation} G_{\mu \nu}
\,=\, \left( \begin{array}{ccccccc}
0&\frac{1}{2}&&&&-\frac{i}{4}\bar{q}_{1}&0\\
\frac{1}{2}&0&&&&\frac{i}{4}q_{1}&0
\\ &&\ddots&&&\vdots&\vdots
\\ &&&0&\frac{1}{2}&-\frac{i}{4}\bar{q}_{r}&0\\
&&&\frac{1}{2}&0&\frac{i}{4}q_{r}&0\\
-\frac{i}{4}\bar{q}_{1}&\frac{i}{4}q_{1}&\ldots&-\frac{i}{4}\bar{q}_{r}&\frac{i}{4}q_{r}&\beta^{2}&\frac{1}{2}\\ 0&0&\ldots&0&0&\frac{1}{2}&0
\end{array} \right) \, , \label{w28a} \end{equation} \vspace{.5cm}
\begin{equation} G^{\mu \nu} \,=\, \left( \begin{array}{ccccccc}
0&2&&&&&-iq_{1} \\ 2&0&&&&&i\bar{q}_{1} \\ &&\ddots&&&&\vdots
\\ &&&0&2&&-iq_{r} \\ &&&2&0&&i\bar{q}_{r} \\ &&&&&0&2 \\
-iq_{1}&i\bar{q}_{1}&\ldots&-iq_{r}&i\bar{q }_{r}&2& \gamma -4\beta^{2}
\end{array} \right) \, , \label{w28b} \end{equation} where the notation
$\gamma \,=\, \sum_{i} q_{i} \bar{q}_{i}$ has been used.  The
antisymmetric field then is found to be \begin{equation}
 B_{q_{i} \bar{q}_{j}} \;=\; \frac{i}{2} u \delta_{ij} \, .\label{w28c}
\end{equation} \par The line element for this case is then written as
\begin{equation}
 ds^2 \;=\; \sum_{i} dq_{i}d\bar{q}_{i}- \left( dv - \frac{i}{2}
 \sum_{i} ( q_{i}d\bar{q}_{i}-\bar{q}_{i}dq_{i}) \right) du + \beta^{2}
 du^2 \, . \label{w29} \end{equation} Introducing polar coordinates
$q_{i}=R_{i}e^{i\theta_{i}}$, $\bar{q}_{i}=R_{i}e^{-i\theta_{i}}$ we
find that the metric is $\eta_{\mu \nu}=(-1,+1,\ldots,+1)$ in the base
\begin{eqnarray}
  e^0 \;&=&\; \frac{1}{2 \beta}(dv-\sum_{i} R_{i}^{2} d\theta_{i}) \, ,
  \nonumber \\ e^{2i-1} \;&=&\; dR_{i} \, , \nonumber \\ e^{2i} \;&=&\;
  R_{i} d\theta_{i} \, , \nonumber \\ e^{D-1} \;&=&\; \beta du - e^{0}
  \, , \label{w30} \end{eqnarray} where $i=1, \ldots ,r$.  The
non-vanishing components of the Ricci tensor in the base (\ref{w30})
are then \begin{equation} {R^{D-1}}_{D-1} \,=\, -{R^{0}}_{0} \,=\,
-{R^{0}}_{D-1} \,=\, \frac{D-2}{4\beta^{2}} \, , \label{w31}
\end{equation} whereas the antisymmetric 3-form field is given by
\begin{equation}
  H \,=\, dB \,=\, \frac{1}{\beta} \sum_{i=1}^{r} \left(e^{D-1} \wedge
  e^{2i-1} \wedge e^{2i} \,+\, e^{0} \wedge e^{2i-1} \wedge e^{2i}
  \right) \, . \label{w32} \end{equation} It is straightforward then to
verify that the vanishing of the one-loop beta functions in
eqs.(\ref{w23}) gives $c=D$.  \par One can understand this result
perturbatively by considering the loop diagrams that contribute to the
beta function.  Since the propagator of the $u$ field vanishes due to
$G^{uu}=G^{D-1 \, D-1}=0$, the only diagrams that can be drawn are
one-loop diagrams with two external $u$ lines and $q_{i}$,
$\bar{q}_{i}$ in the internal lines.  However, the interaction terms in
(29) give opposite contributions to these
 diagrams, in exactly the same way discussed by Nappi and Witten
 \cite{1,9} leading to $c=D$ identically in perturbation theory.  \par
The same result can be obtained non-perturbatively as follows
\cite{4,5,10}:  The Noether currents associated with the WZW model
(\ref{w4}) have the following OPE \begin{equation}
  J_{I}(z) J_{J} (w) \,=\, -\frac{k\, \Omega_{IJ}}{2 (z-w)^2} +
  {f_{IJ}}^{K} \frac{J_{K}(w)}{(z-w)}  + reg.  \end{equation} Let us
define the bilinear in the currents energy-momentum tensor
\begin{equation}
 T(z) \,=\, L^{IJ} : \!\! J_{I} J_{J} \!\! : \!\! (z) \label{star1} \,
 .  \end{equation} The condition for the $J_{I}$'s to be primary fields
of weight 1, is written as \begin{equation}
 T(z) J_{I} (w) \,=\, \frac{J_{I} (w)}{(z-w)^2} + \frac{ \partial J_{I}
 (w)}{(z-w)} + reg.  \label{star2} \end{equation} and by using
eq.(\ref{star1}) we get \cite{4,5} \begin{eqnarray}
   L^{IJ} {f_{KJ}}^{L} \,+\, L^{JL} {f_{KJ}}^{I} \;&=&\; 0 \, ,
   \label{vir1} \\ 2 L^{IJ}\Omega_{KJ} + L^{MN}{f_{KM}}^{L}
 {f_{LN}}^{I}\;&=&\; {\delta^{I}}_{K} \, . \label{vir2} \end{eqnarray}
Thus $L^{IJ}$ is an invariant symmetric tensor as follows from
eq.(\ref{vir1}) and employing the latter in eq.(\ref{vir2}) we get
\begin{equation}
      L^{IJ} ( 2 \Omega_{KJ} + \kappa_{KJ} ) \;=\; {\delta^{I}}_{K}
      \label{maarja} \, , \end{equation} where $\kappa_{KJ} =
{f_{KM}}^{L} {f_{JL}}^{M}$ is the Killing form.  Then it follows that
$L^{IJ}$ is the inverse of the matrix $(2\Omega_{KJ}+\kappa_{KJ})$ and
in our case one finds that \begin{equation}
 L^{IJ} \,=\, \frac{1}{2a} \left( \begin{array}{ccccccc} 0&1&&&&&
 \\ 1&0&&&&& \\ &&\ddots&&&& \\ &&&0&1&& \\ &&&1&0&& \\ &&&&&0&-1
 \\ &&&&&-1& -\frac{b+r}{a} \end{array} \right) \, . \label{w36}
\end{equation} The central charge is given by \begin{equation}
  c \;=\; 2 L^{IJ} \Omega_{IJ} \;=\; D \, , \label{centr1}
\end{equation} in accordance with the results in \cite{4} since one may
look upon the Heisenberg algebra {\cal H}$_{D}$ as the double $U(1)$
extention of the algebra \begin{equation}
	[ \an_{i} , {\an_{j}}^{\dagger} ] \;=\; 0 \, .  \end{equation}
\par The above construction is a direct generalization of Nappi and
Witten's result \cite{1}, which is recovered in the case $D=4$.  The
space-time described by the metric (\ref{w29}) is homogeneous, it has
$2(D-1)$ space-like Killing vectors and a null one, a total of $2D-1$.
It describes a plane-wave type space-time with flat wave surfaces
($u,v= const.$) since the coset space {\cal H}$_{D}/U(1)\! \times \!
U(1)$ is isomorphic to $\notC^{r}$.  It has one time-like direction
 (Lorentz signature $D-2$) and thus it can be considered as a
$D$-dimensional string background which can replace Minkowski
space-time.  \vspace{7mm}

We would like to thank P. de Boer for his help in the computer programm
and N. Obers for discussions.


\newpage \begin{thebibliography}{99} \bibitem{1} C.R. Nappi and E.
Witten, Phys. Rev. Lett. 71 (1993)3751.  \bibitem{2} D.I. Olive, E.
Rabinovici and A. Schwimmer, Phys. Lett. B 321 (1994)361.  \bibitem{3}
K. Sfetsos, {\em ``Exact string backgrounds from WZW models based on
non-semi-simple groups"}, preprint THU-93/31, hep-th/9311093; {\em
``Gauged WZW models and non-abelian duality"}, preprint THU-94/1,
hep-th/9402031.  \bibitem{4} J.M. Figueroa-O'Farrill and S. Stanciu,
{\em ``Nonsemisimple Sugawara constructions"}, preprint QMW-PH-94-2,
hep-th/9402035.  \bibitem{5} N. Mohammedi, {\em ``On bosonic and
supersymmetric current algebras for non-semi-simple groups"}, preprint
BONN-HE-93-51, hep-th/9312182.  \bibitem{6} E. Kiritsis and C. Kounnas,
Phys. Lett. B 320 (1994)264.  \bibitem{7} H.W. Brinkmann, Proc. Nat.
Acad. Sci. (U.S.) 9 (1923) 1; \\
 N. Rosen, Phys. Z. Sowjetunion, 12.4 (1937)366; \\ J. Ehlers and W.
 Kundt, in {\em Gravitation: an introduction to current research}, ed.
 L. Witten, J.W., N.Y. 1962.  \bibitem{8} R. G\"uven, Phys. Lett. B 191
(1987)275; \\
 D. Amati and C. Klim\v{c}ik, Phys. Lett. B 219 (1989)443; \\ G.T.
 Horowitz and A.R. Steif, Phys. Rev. Lett. 64 (1990)260; \\ A.A.
 Tseytlin, Nucl. Phys. B 390 (1993)153; Phys. Rev. D 47 (1993)3421.
\bibitem{9} C.R. Nappi, Phys. Rev. D 21 (1980)418; \\ E. Witten,
Commun. Math. Phys. 92 (1984)455.  \bibitem{10} M.B. Halpern and E.
Kiritsis, Mod. Phys. Lett. A 4 (1989)1373; Erratum Mod. Phys. Lett. A 4
(1989)1797.  \end{thebibliography}
\end{document}